\documentclass[onecolumn,showkeys,preprintnumbers,aps,a4paper,amssymb,prd,superscriptaddress,nofootinbib]{revtex4-2}
\usepackage{comment}
\usepackage{graphicx}
\usepackage{epsf}
\usepackage{bm}
\usepackage{amsmath}
\usepackage{amsfonts}
\usepackage{amssymb}
\usepackage{epstopdf}
\usepackage{color}
\usepackage[dvipsnames]{xcolor}
\usepackage{verbatim}
\usepackage{multirow}
\usepackage{soul}
\usepackage{physics}
\usepackage{bm}

\usepackage[width=0.00cm, height=0.00cm, left=1.50cm, right=1.50cm, top=2.00cm, bottom=2.00cm]{geometry}
\usepackage{microtype}
\usepackage{lmodern}

\usepackage[colorlinks = true,
            linkcolor = teal,
            urlcolor  = teal,
            citecolor = teal,
            anchorcolor = blue]{hyperref}
\usepackage[capitalize]{cleveref}
\usepackage[normalem]{ulem}
\usepackage{enumitem}
\usepackage{booktabs}

\usepackage{lipsum}

\makeatletter\let\expandableinput\@@input\makeatother

\begin{document}

\begin{center}
		\vspace{0.4cm} {\large{\bf Dynamical Dark Energy Signatures from a New Transition $Om(z)$ Parametrization in Flat FLRW Cosmology}} \\
		\vspace{0.4cm}
		\normalsize{ Manish Yadav$^1$, Archana Dixit$^2$, Anirudh Pradhan $^3$, M. S. Barak$^4$ }\\
		\vspace{5mm}
		\normalsize{$^{1,4}$ Department of Mathematics, Indira Gandhi University, Meerpur, Haryana 122502, India\\
		\normalsize{$^{2 }$ Department of Mathematics, Gurugram University Gurugram, Harayana, India}.} \\

        	\normalsize{$^{3 }$ Centre for Cosmology, Astrophysics and Space Science (CCASS), GLA University, Mathura, Uttar Pradesh, India}\\ 
		\vspace{2mm}
			$^1$Email address: manish.math.rs@igu.ac.in\\
		    $^2$Email address: archana.ibs.maths@gmail.com\\
                $^3$Email address: pradhan.anirudh@gmail.com\\
             $^4$Email address: ms$_{-}$barak@igu.ac.in\\
\end{center}

 
\maketitle
{\noindent {\bf Abstract.}

We investigate a cosmic scenario using a new transition parameterization of the $Om(z)$ diagnostic, $Om(z) = \frac{z^l}{(1+z)^m}$, in the spatially flat Friedmann Lemaître Robertson-Walker (FLRW) framework. Using observational datasets such as Observational Hubble Data (OHD), Pantheon Plus (PP), and SH0ES, we analyze the evolution of the $Om(z)$ function to probe deviations from the standard $\Lambda$CDM model and constrain free parameter space {$H_0$, l, m } using Markov Chain Monte Carlo (MCMC) analysis with the emcee sampler. Our analysis reveals a clear transition in the slope of $Om(z)$ from negative to positive at transition redshift values $z_t \approx 1.41$, $0.65$, and $0.33$ for the OHD, OHD+PP, and OHD+PP$\&$SH0ES datasets, respectively. This behavior suggests a dynamical evolution of dark energy, indicating a transition from a quintessence-like phase to a phantom regime. From the combined OHD+PP$\&$SH0ES dataset, we obtain a best-fit value of the Hubble constant \( H_0 = 73.01 \pm 0.36 \, \mathrm{km\,s^{-1}\,Mpc^{-1}} \), which is consistent with the SH0ES calibration and supports the viability of our model. Additionally, our analysis indicates that the current age of the Universe is approximately $13 \sim 14$ Gyr from all available combinations of datasets, which is consistent with observational expectations. Further, we find that the deceleration-to-acceleration transition, which marks the beginning of cosmic acceleration, is inferred to occur within the redshift interval $z_t \in [0.5, 0.8]$, highlighting the emergence of dark energy as the dominant component in the Universe's recent expansion history. Our transition $Om(z)$  parameterization captured progressive cosmological changes and enabled seamless interpolation over cosmic epochs.
\smallskip 


\section{Introduction}

One of the most significant discoveries in modern cosmology came at the end of the 20th century, when two independent research groups revealed that the universe is undergoing accelerated expansion, this phenomenon was first indicated by observations of Type Ia supernovae (SN Ia) \cite{ref1,ref2,ref3} and later corroborated by a range of independent astrophysical probes, including measurements of the cosmic microwave background (CMB) \cite{ref4,ref5,ref6},  large-scale structure surveys \cite{ref7} and the detection of baryon acoustic oscillation (BAO) peaks \cite{ref8,ref9,ref10,ref11}. The discovery of this accelerated expansion raised fundamental questions regarding the nature of the component responsible for this effect. To account for this, a mysterious form of energy with negative pressure, known as dark energy (DE), was proposed by the cosmological community. Understanding the nature of dark energy through its EoS ( \( w = \frac{p}{\rho} \), where $p$ is pressure and $\rho$ is the density of the DE of the universe) has become a central focus in modern cosmology. This parameter is crucial for characterizing the dynamic behavior of dark energy. While many cosmologist researchers used the cosmological constant $(w = -1)$ in our research owing to its simplicity and  strong agreement with a wide range of astronomical observational data, recent studies have highlighted several theoretical inconsistencies associated with the cosmological constant itself. One of the most significant concerns is the fine-tuning problem \cite{ref12}with a cosmological constant. Observations reveal that the cosmological constant possesses an exceedingly small value that is sufficient to explain the observed acceleration of the cosmos. However, predictions made using quantum field theory indicate a vastly larger value for $\Lambda$, with discrepancies arising from a staggering factor of $10^{120}$. However, the coincidence problem \cite{ref13} highlights the puzzlingly similar orders of magnitude of $\Lambda$ and the density of ordinary baryonic matter in the present universe. Because of these unresolved issues, the possibility that the cosmic EoS varies with time has been explored in a variety of dynamical dark energy models, including quintessence \cite{ref14}, K-essence \cite{ref15,ref16}, and phantom fields \cite{ref17,ref18}. Several studies in the cosmology literature are essential to develop robust methods to determine whether the EoS is time-dependent \cite{ref19,ref20,ref21,ref22}. This distinction is key to improving our understanding of the fundamental nature of dark energy and to guiding the development of more accurate cosmological models.\\

In cosmology, the $Om(z)$ diagnostics have been widely used to classify different types of dark energy (DE)  models and identify potential departures from the mainstream standard model framework. The method was first introduced by Sahni et al.\cite{ref23} as a model-independent tool for testing the nature of DE based on the $Om(z)$ diagnostic slope, and has since been refined and expanded.  In a subsequent study, Sahni et al. \cite{ref24} used an improved diagnostic method to assess the $\Lambda$CDM hypothesis along with BAO data to provide model-independent evidence for evolving dark energy. Their results raise the possibility of departures from the standard model and hint at the dynamic characteristics of the DE component. Subsequent work by Ding et al. \cite{ref25} combined data from gravitational lensing, Type Ia supernovae, and BAOs to further investigate DE evolution using a diagnostic method. Zheng et al. \cite{ref26} used observational data to study both the $Om$ and related diagnostics in this context. Seikel et al. \cite{ref27} assessed the consistency of the $\Lambda$CDM model through a non-parametric reconstruction of $Om(z)$ using Gaussian processes and data from the SNe Ia and Hubble parameter measurements. In recent years, several parameterizations of the $Om(z)$ diagnostic have been introduced to better explore potential departures from $\Lambda$CDM and examine the dynamical behavior of dark energy.  Myrzakulov et al., \cite{ref28}  adopted a power-law parameterization approach of $Om(z)$ diagnostic as, $Om(z) = \alpha (1 + z)^n$, where $\alpha$ and $n$ are parameters that allow for deviations from standard cosmology. In a later study, Myrzakulov et al., \cite{ref29} employed a logarithmic form of the $Om(z)$ parameterization, offering another approach to capture deviations from the $\Lambda$CDM framework. Another frequently studied form is the Chevalier–Polarski–Linder (CPL) parameterization, which models $Om(z)$ analogously to evolving EoS models \cite{ref30}. \\

In this study, we propose a new transition parameterization of the $Om(z)$ diagnostic, defined as $Om(z)  =  \frac{z^l}{(1 + z)^m}$, where $l$ and $m$ are free parameters. This form is physically motivated and offers flexibility in capturing gradual cosmological changes, and transition parameterization allows for smooth interpolation across cosmic epochs. It provides an efficient method to probe potential deviations from $\Lambda$CDM using a minimal number of parameters, building upon and extending previous studies based on the $Om(z)$ diagnostic. \\

In the existing literature, a number of studies have demonstrated a single dominant trend in the behavior of the $Om(z)$ diagnostic, typically exhibiting either a consistently positive or consistently negative slope. This limitation restricts their ability to capture the full complexity of dark energy dynamics across cosmic time. Motivated by the need for a more comprehensive understanding, our work introduces a cosmological model that is both flexible and dynamically rich and capable of reproducing both trends—namely, a positive slope (indicative of phantom-like behavior) and a negative slope (characteristic of quintessence-like behavior). This dual capability allows our model to describe a possible transition in the nature of dark energy within a unified theoretical framework, offering deeper insight into its evolving behavior across different redshift regimes. \\

The motivation for adopting the $Om(z)$ diagnostic form is its ability to naturally describe a smooth cosmological transition while remaining mathematically simple and physically meaningful. The power-law dependence on redshift ensures regular behavior at both low and high redshifts, enabling continuous interpolation across different cosmic epochs without introducing artificial features. The presence of two independent parameters allows the model to capture a wide range of possible evolutionary behaviors of $Om(z)$, including both increasing and decreasing trends with a redshift. Such flexibility is crucial for probing dynamical dark energy scenarios, as different trends in $Om(z)$ are directly linked to phantom-like and quintessence-like behaviors. Unlike many existing $Om(z)$ parameterizations, which are restricted to a single monotonic trend, the present form can accommodate a possible transition in the nature of dark energy within a unified framework. At the same time, the parametrization reduces to behavior consistent with the standard $\Lambda$CDM model for suitable parameter choices, making it an efficient and robust diagnostic tool to test deviations from $\Lambda$CDM using observational data.\\

The present manuscript is organized into several key sections. In Section II, we introduce the FLRW universe and propose a new transition parameterization for the $Om(z)$ diagnostic, providing a comprehensive overview. Based on this, we constructed our derived cosmological model. In Section III, we introduce the latest OHD, and PP\&SH0ES data sets and the methodology employed in our analysis. Section IV  presents the results of the analysis and discusses the key findings. Finally, Section V provides a summary and conclusion.

\section{MODEL}
\label{sec:datasets}

The study of the universe’s large-scale dynamics typically relies on cosmological principles (spatial homogeneity and isotropy at sufficiently large scales). Although the direct verification of this principle across all epochs and scales remains elusive,  it is strongly supported by a comprehensive and broader range of observational data. Under this framework, the spacetime geometry of the universe is described by the FLRW metric, which is expressed as
\begin{equation}
\label{flrw}
ds^2 = -c^2 dt^2 + a(t)^2 \left[ \frac{dr^2}{1 - \kappa r^2} + r^2 \left( d\theta^2 + \sin^2\theta \, d\phi^2 \right) \right],
\end{equation}

where $a(t)$ is the cosmic scale factor change with cosmic time and $\kappa$ represents the curvature scalar, which represents open, flat, and closed universes with $\kappa < 0$, $\kappa = 0$, and $\kappa > 0$, respectively. In modern observational cosmology—particularly with cosmic data from  the CMB,  BAO, and SN Ia - there is compelling evidence that the universe is spatially flat. Accordingly, we adopt  $\kappa = 0,$ and $ 8\pi G = c = 1$. Under this condition, the first and second Friedmann equation reduces to:

\begin{equation}
\label{Frd1}
3 H^2 = \rho_{tot}
\end{equation}

\begin{equation}
\label{Frd2}
2\dot{H} + 3\mathcal H^2 = -p_{tot},
\end{equation}
where $ {H} = \dfrac{\dot{a}}{a}$,  and $\dot{H} = \dfrac{\ddot{a}}{a}$ represent the Hubble parameter and the rate of change of the Hubble parameter with respect to cosmic time $t$, respectively. Here $\rho_{tot}$ and $p_{tot}$ are the total energy density and total pressure of the fluid in the universe, respectively. \\

The $Om(z)$ diagnostic is a powerful and insightful tool introduced as a null test for dark energy (DE) models \cite{ref23}. The beauty of $Om(z)$ is directly connected to the Hubble parameter $H(z)$ as well as the redshift $z$, which is an observable quantity that can be measured independently from various astrophysical and cosmological data, such as SNeIa and BAO. The $Om(z)$ provides a practical and observationally relevant approach for exploring the characteristics of the dark energy. A key strength of the $Om(z)$ diagnostic is its capacity to distinguish between a cosmological constant—representing the simplest dark energy model characterized by a constant energy density, and more intricate dynamic dark energy scenarios, where the energy density varies over time. This capability makes $Om(z)$ a valuable tool in cosmology, enhancing our understanding of the universe's accelerating expansion and the mechanisms underlying dark energy.\\

The $Om(z)$ diagnostic serves as a basis for distinguishing between three major forms of dark energy, which are presented in Ref.\cite{ref23}. If $Om(z)$ is constant across all redshifts, it strongly favours the interpretation that dark energy behaves as a cosmological constant, often referred to as $\Lambda$. On the other hand, if $Om(z)$ changes with the redshift, it suggests the existence of dynamical dark energy, which may arise from new physics beyond the standard cosmological constant framework. The behaviour of the slope $Om(z)$ as a function of redshift provides even deeper insights into the type of dynamical model at play. Specifically, a positive slope $Om(z)$ corresponds to a so-called "phantom" dark energy phase, where the equation of state parameter $w$ (which relates pressure and energy density) is less than $-1$. Phantom models predict unusual cosmic dynamics, including the possibility of a future "Big Rip" singularity. Conversely, a negative slope in $Om(z)$ points toward quintessence models of dark energy, where it $w$ lies between $-1$ and $-\frac{1}{3}$, and describes a more conventional scalar field slowly rolling down its potential. Therefore, analyzing the slope $Om(z)$ allows cosmologists to not only detect deviations from the cosmological constant, but also to classify the dynamic nature of dark energy.\\

Several studies have used reconstructed observational data and \( Om(z) \) as a diagnostic tool to assess the consistency of the standard \( \Lambda \)CDM model. For example, \( Om(z) \) has been used in publications like \cite{ref23,ref24,ref25,ref26,ref27,ref28,ref29,ref30,ref31,ref32} to investigate whether the cosmological constant hypothesis is true under various reconstruction methods and data sets. By complementing existing cosmological probes and offering crucial cross-checks, these initiatives help strengthen our understanding of cosmic acceleration and underlying physics. We propose a novel parameterization of $ Om(z) $ as a function of the redshift, motivated by these theoretical considerations and empirical findings, given by 

\begin{equation}
Om(z)  =  \frac{z^l}{(1 + z)^m}
\label{eq4}
\end{equation}

Here, \( l \) and \( m \) are the free parameters of the model that vary with the redshift. The behavior of $ Om(z)$ depends on several conditions of l and m, however three key conditions involving these parameters can be clearly identified in Fig. \ref{fig1}. 
\begin{itemize}

 \item If \( l > m \), \( Om(z) \) exhibits a positive slope, indicating the presence of a phantom regime with the equation of state \( w < -1 \). 

\item  If \( l < m \), the slope of \( Om(z) \) can be both positive and negative, suggesting a transition between phantom and quintessence behaviors. 

\item  In the special case where \( l = 0 \) and \( m > 0 \), \( Om(z) \) exhibits a negative slope, corresponding to a quintessence region characterized by \( w > -1 \).
\end{itemize}

\begin{figure}[hbt!]
     \centering
     \includegraphics[width=0.7\linewidth]{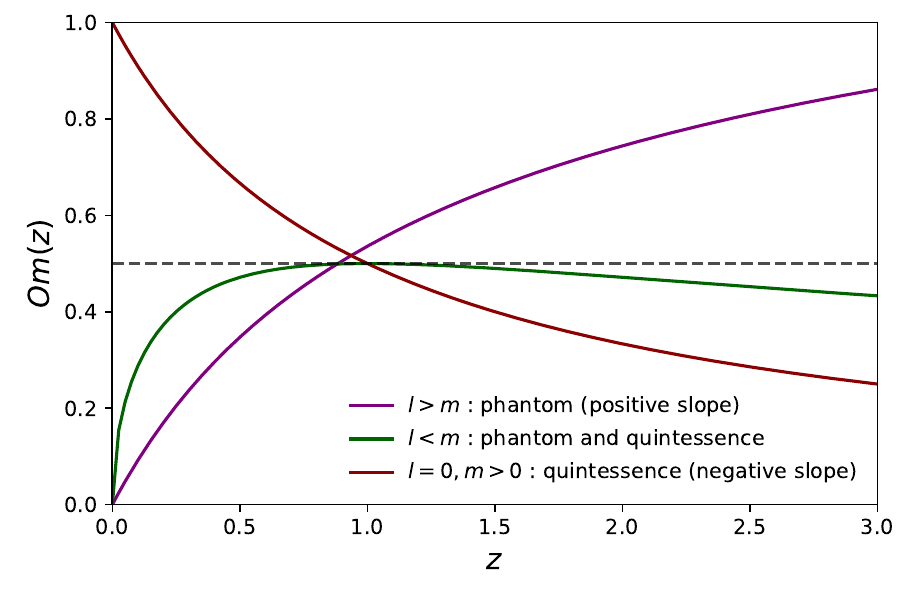}
     \caption{Evolution of the diagnostic function 	$Om(z)$ for different dark-energy models. The purple curve ($l>m$) corresponds to a phantom model with a positive slope, the green curve ($l<m$) shows mixed phantom and quintessence behavior, and the red curve ($l=0,m>0$) represents a quintessence model with a negative slope. The dashed horizontal line marks a constant reference value of $Om(z)$.}
     \label{fig1}
 \end{figure}

We calculated the Hubble parameter in terms of \( Om \) diagnosis and z with the help of $\text{Om}(z) = \frac{H^2(z)/H_0^2 - 1}{(1+z)^3 - 1}$ is given by 

\begin{equation}
H(z) = H_0 \sqrt{1 + Om(z) \left( (1 + z)^3 - 1 \right)}
\label{eq5}
\end{equation}
 From Eqs.\ref{eq4} and \ref{eq5}, we obtain the governing equation for the Hubble parameter as follows:
\begin{equation}
H(z) = H_0 \sqrt{1 + \left( \frac{z^l}{(1 + z)^m} \right) \left( (1 + z)^3 - 1 \right)}
\label{eq6}
\end{equation}
where $H_0$ is present Hubble constant value at $z = 0$\\
From Eq.\ref{Frd1}, the total energy density of the universe in terms of redshift is expressed as 

\begin{equation}
\rho_{\text{total}}(z) = 3 H_0^2 \left[ 1 + \left( \frac{z^l}{(1+z)^m} \right) \left( (1+z)^3 - 1 \right) \right]
\label{eq7}
\end{equation}

Now, we evaluate total EoS parameter and deceleration parameter in terms of redshift as

\begin{equation}
w_{\text{tot}}(z) = -1 + \frac{(1+z)}{3 \left[1 + \frac{z^l}{(1+z)^m} \left((1+z)^3 - 1\right)\right]} 
\times 
\left\{
\left[
l z^{l-1} (1+z)^{-m} - m z^l (1+z)^{-m-1}
\right] \left((1+z)^3 - 1\right)
+ \frac{z^l}{(1+z)^m} \cdot 3(1+z)^2
\right\}
\label{eq8}
\end{equation}

\begin{equation}
q(z) = -1 + \frac{(1+z)}{2 \left[1 + \frac{z^l}{(1+z)^m} \left((1+z)^3 - 1\right)\right]} 
\times 
\left\{
\left[
l z^{l-1} (1+z)^{-m} - m z^l (1+z)^{-m-1}
\right] \left((1+z)^3 - 1\right)
+ \frac{z^l}{(1+z)^m} \cdot 3(1+z)^2
\right\}
\label{eq9}
\end{equation}

 The essential characteristics of the \( Om \) diagnosis are captured by these  parameters, which are  adaptable enough to cover a broad spectrum of dark energy behavior, from a cosmological constant to different dynamical situations. Our goal was to verify the validity of our theoretical models using the latest observational data, such as OHD, PP, and PP\&SH0ES.

\section{DATASETS AND METHODOLOGY}

\textbf{Observational Hubble Data }: In this segment, we make use of 33 independent measurements of the Hubble parameter $H(z)$, covering the redshift interval $0.07 \leq z \leq 1.965$ \cite{ref33,ref34,ref35,ref36,ref36a,ref37,ref38,ref39,ref40}, obtained through the Observational Hubble data (OHD) approach. This method relies on estimating the differential ages of passively evolving galaxies, allowing for a model-independent determination of the expansion rate of the universe by relating the Hubble parameter to the redshift and cosmic time via the following relation:

\begin{equation}
H(z) = -\frac{1}{1+z} \frac{dz}{dt},	
\end{equation}

as originally proposed in foundational studies~\cite{ref41}. In our analysis, we employed these measurements to constrain the parameters of the proposed cosmological scenario. The theoretical predictions of the Hubble value from our model-dependent approach showed good agreement with the model-independent OHD and $\Lambda$CDM model across the redshift range, as illustrated in Fig.\ref{fig2}.\\

\begin{figure}[hbt!]
    \centering
    \includegraphics[width=1\linewidth]{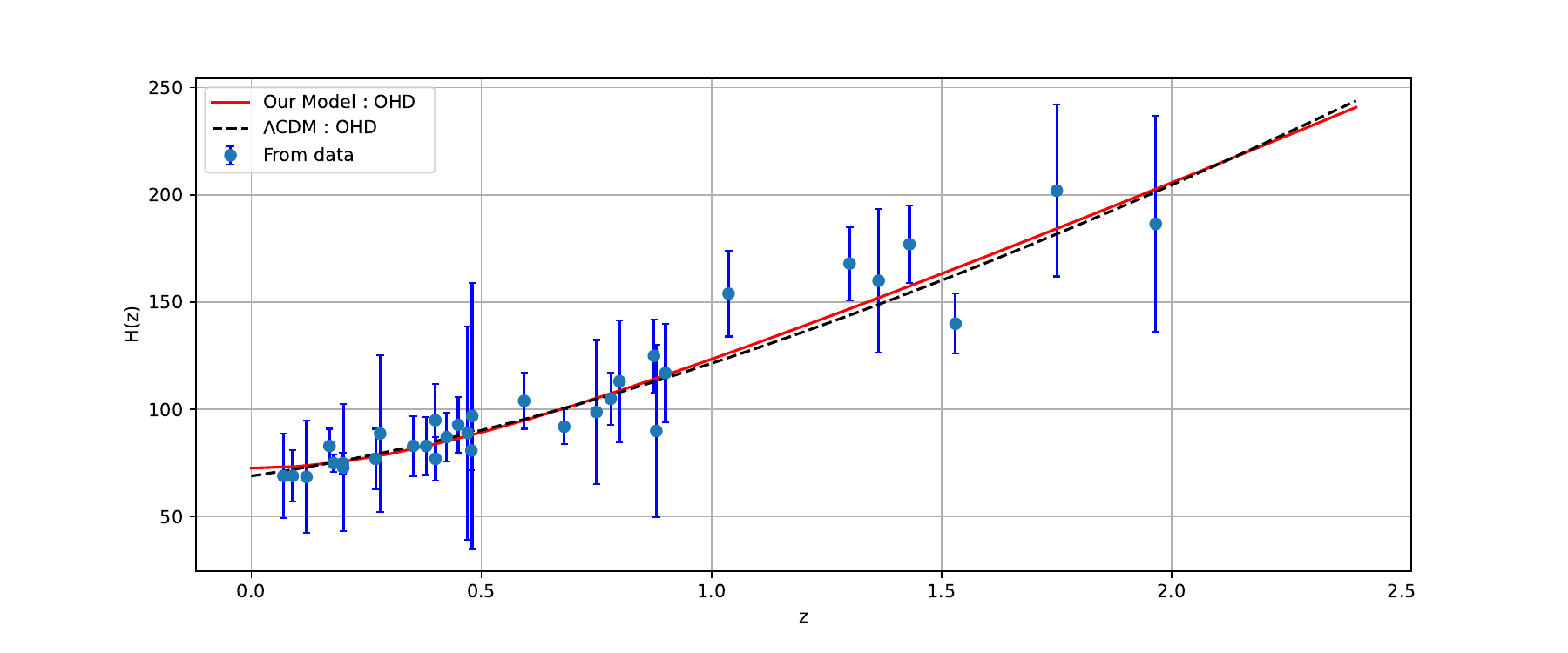}
    \caption{The red curve for our model and black dotted line for $\Lambda$CDM model with error bar (blue colour).}
    \label{fig2}
\end{figure}

 \textbf{PantheonPlus\&SH0ES}: Type Ia supernovae serve as valuable tools for measuring distance moduli, which help constrain the uncalibrated luminosity distance multiplied by the Hubble constant, $H_0 d_L(z)$. For a supernova observed at redshift $z$, the distance modulus is defined as 
 
\begin{equation}
\mu = 5 \log_{10} d_L(z) + 25,
\end{equation}

where $\mu = m_B - M_B$ represents the difference between the observed apparent magnitude and the intrinsic absolute magnitude.  The luminosity distance, $d_L$, is a key cosmological probe that traces the expansion history of the universe. In the case of a spatially flat universe, the luminosity distance is expressed as 

\begin{equation}
d_L(z) = (1 + z) \times D_H(z),
\end{equation}

where $D_H(z)$ denotes the comoving Hubble distance, and can be expressed as $D_H(z) = \int_0^z \frac{dz'}{H(z')}$\\

In this study, we utilized the Pantheon compilation sample of Type Ia supernova data from references \cite{ref42}, and the PantheonPlus dataset consists of 1701 light curves corresponding to 1550 distinct SNe Ia events, spanning a redshift range of \( z \in [0.001, 2.26]\). This dataset calibrates the Type Ia supernova magnitude using additional cepheid hot distances \cite{ref43,ref43a}. We used the label of Pantheon data as PP and Pantheon Plus SH0ES data as PP\&SH0ES. As shown in Fig.\ref{fig3}, our model-dependent curve shows good agreement with the model-independent PP\&SH0ES observation data as well as $\Lambda$CDM model curve across the redshift range.

\begin{figure}[hbt!]
    \centering
    \includegraphics[width=1\linewidth]{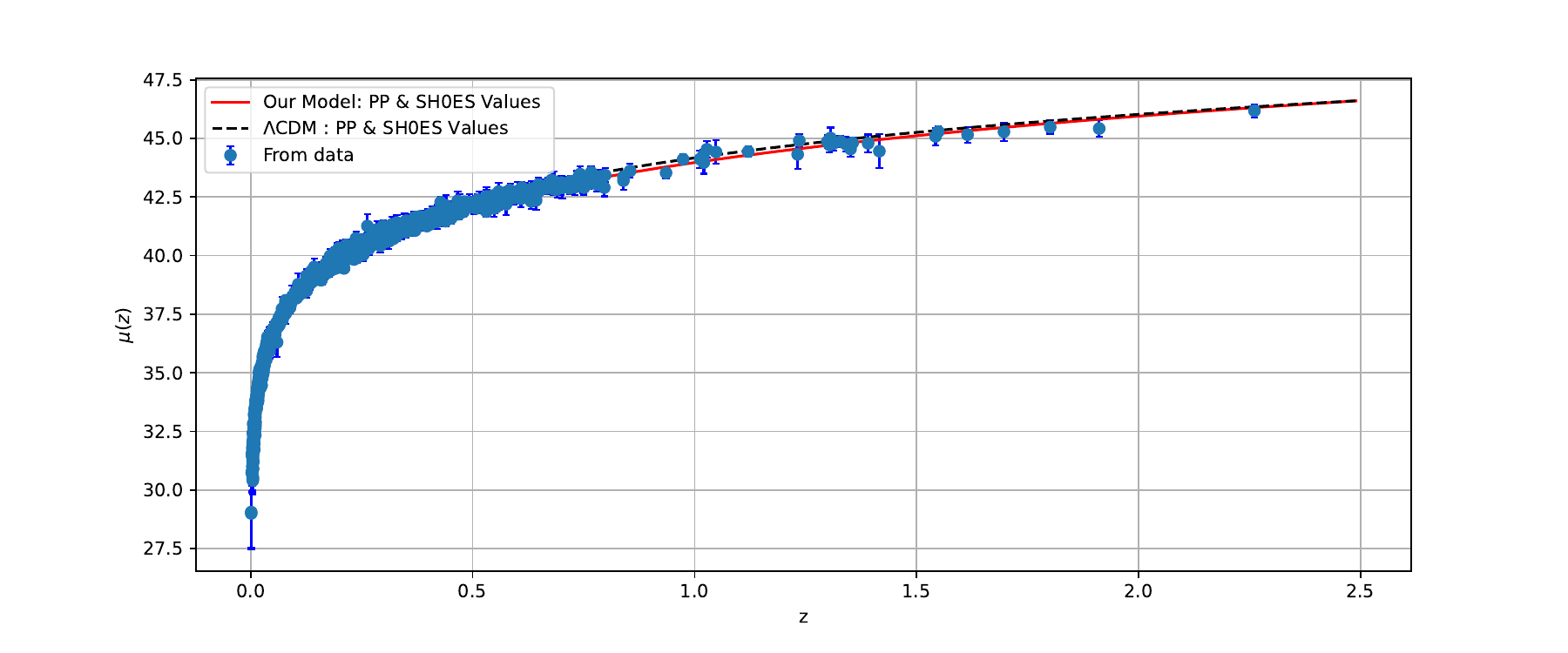}
    \caption{The 2D plot shows the distance modulus $µ(z)$ for our model (red line) and the $\Lambda$CDM model (black dotted line) with corresponding blue colour error bars.}
    \label{fig3}
\end{figure}

To statistically evaluate the consistency between the theoretical model and  observational data, we quantified the goodness-of-fit of our model by defining a statistical $\chi^2$ function based on the joint analysis of OHD, PP, and PP\&SH0ES data,

\begin{equation}
\chi^2_{\text{joint}} =  \chi^2_{\text{OHD}} + \chi^2_{\text{PP}} + \chi^2_{\text{PP\&SH0ES}},
\end{equation}

where,

\begin{equation}
\chi^2_{\text{OHD}} = \sum_{i=1}^{33} \frac{\left[ d^{obs}(z_i) - d^{th}(z_i) \right]^2}{\sigma^2_{d^{obs}(z_i)}},
\end{equation}

\begin{equation}
\chi^2_{\text{PP}} = \sum_{i,j}^{1048} \Delta\mu_i \left(C_{\text{stat + syst}}^{-1}\right)_{ij} \Delta\mu_j.
\end{equation}

\begin{equation}
\chi^2_{\text{PP\&SH0ES}} = \sum_{i,j}^{1701} \Delta\mu'_i \left(C_{\text{stat + syst}}'^{-1}\right)_{ij} \Delta\mu'_j.
\end{equation}

In this analysis, $ d^{obs}(z_i)$ and $ d^{th}(z_i)$ denote the observed and model-predicted values of the Hubble parameter at redshift $z_i$, respectively, and $\sigma_{{d^{obs}(z_i)}}$ represents the associated observational uncertainty, as provided in Ref. Here, $\Delta \mu_i = \mu^{\text{th}}_i - \mu^{\text{obs}}_i$ represents the deviation between the theoretical and observed distance modulus values. The matrix $C_{\text{stat + syst}}^{-1}$ is the inverse of the covariance matrix corresponding to the Pantheon dataset, which accounts for statistical and systematic ( $C_{\text{stat + syst}} = C_\text{stat} + C_\text{syst}$) correlations between supernova measurements. Similarly, $\Delta \mu'_i = \mu^{\text{th}}_i - \mu^{\text{obs}}_i$ represents the deviation between the theoretical and observed distance modulus values. Matrix $C_{\text{stat + syst}}^{-1}$ is the inverse of the covariance matrix corresponding to the PP\&SH0ES dataset, which accounts for statistical and systematic ( $C'_{\text{stat + syst}} = C'_\text{stat} + C'_\text{syst}$) correlations between supernova measurements. \\

In our analysis, we utilize the Markov Chain Monte Carlo (MCMC) method to constrain cosmological parameters by analyzing astrophysical observational data, primarily focusing on constraining the free parameter space $(H_0,l, m )$ with corresponding the ranges $H_0 \in [60, 80]$, $l \in [0, 2]$, and $m \in [0, 10]$ respectively. The \texttt{emcee} library \cite{ref44} was employed for parallelized MCMC sampling using 80 walkers and 10000 steps to ensure convergence. By independently and jointly analyzing the  33 observational $H(z)$ data points, and 1701 supernova measurements from the Pantheon+ and SH0ES compilations, we can extract meaningful constraints on the cosmological parameters and better understand the expansion history of the universe.\\

\section{Results and Disussion}

In this study, we constrained the free parameter space ($H_0$, $l$, $m$) using a switching transition model, based on individual or joint combinations of observational datasets such as OHD, PP, and PP\&SH0ES. Our primary objective was to analyze the newly constructed $Om(z)$ diagnostic parameter across all considered datasets. In Section 2, we  discussed the complete behavior of the dark energy under various conditions imposed on the free parameters. However, our derived model exhibits both the phantom and quintessence regimes when applied to the three considered datasets, with this dual behavior emerging specifically under the condition that the free parameter $l<m$.

\begin{figure}[hbt!]
     \centering
     \includegraphics[width=0.6\linewidth]{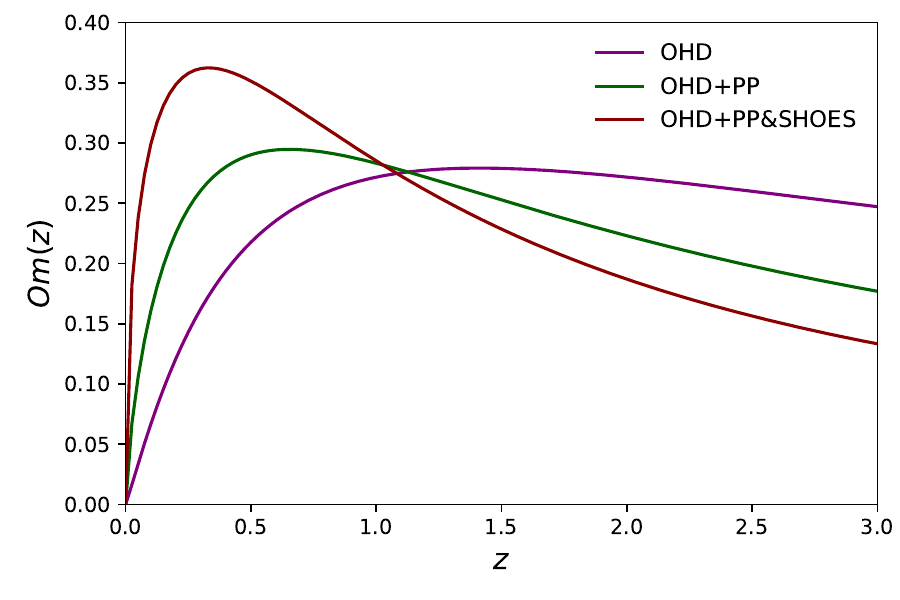}
     \caption{The reconstructed  trajectories (from present to past) of $Om(z)$ diagnostic in our model based on the given datasets.}
     \label{fig4}
 \end{figure}

As shown in Fig.\ref{fig4}, the 2D plot of the $Om(z)$ diagnostic is a function of the redshift, $z$. In this figure, the purple curve, representing the OHD data, shows a positive slope in the range of redshift $0 < z < 1.41$, indicating  phantom-like behavior of dark energy. At $z = 1.41$, the slope of the curve became zero, indicating a transition point. At this redshift, the behavior of the dark energy changes which is referred to as a transition redshift. Beyond this point, the slope becomes negative, which is consistent with  quintessence-like behavior. This analysis highlights a clear transition in the nature of dark energy from  a quintessence to a phantom regime at transition redshift $z = 1.41$. When the PP data were combined with the OHD dataset, a significant shift in the transition redshift is observed, as illustrated by the green curve in the figure. The curve exhibits a positive slope in the redshift range $0 < z < 0.65$, indicating a phantom-like behavior of dark energy in this regime. Beyond the redshift $z = 0.65$, the slope changes, aligning with a quintessence-like behavior. This analysis clearly demonstrates a transition in the nature of dark energy—from  quintessence to phantom occurring at a transition redshift of $z = 0.65$. A further refinement of the analysis is achieved by combining OHD with the PP\&SH0ES dataset. The resulting red curve in the figure reveals yet another significant shift in the transition redshift. Here, the slope is positive for $0 < z < 0.33$, suggesting phantom-like dynamics, while beyond $z = 0.33$, the slope turns negative, signaling a transition to a quintessence-like phase. This confirms a clear and early transition in the nature of dark energy at a critical redshift of $z = 0.33$.\\

\begin{figure}[hbt!]
    \centering
    \includegraphics[width=0.8\linewidth]{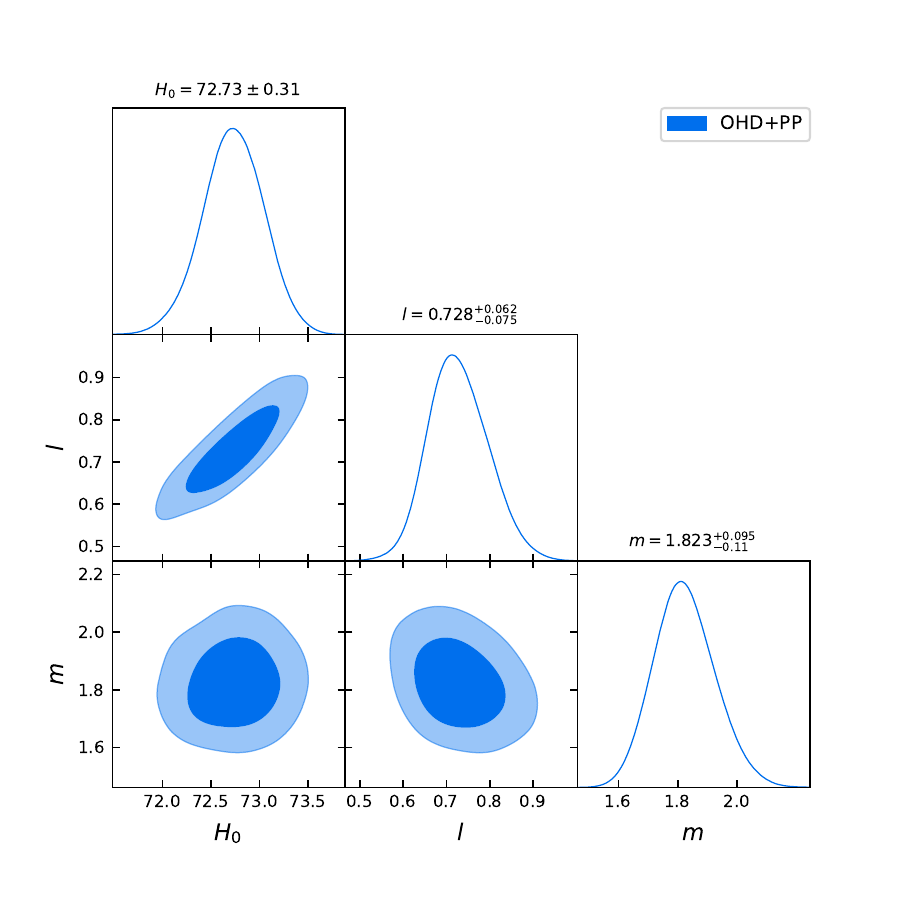}
    \caption{The triangular plot with $1\sigma$ and $2\sigma$ confidence levels for our model based on OHD+PP datasets. }
    \label{fig5}
\end{figure}

We observe that the numerical values of the transition redshift—where the nature of dark energy shifts from a quintessence-like regime to a phantom-like regime  are found to be $z = 1.31$, $z = 0.65$, and $z = 0.33$, corresponding to the OHD, OHD+PP, and OHD+PP\&SH0ES datasets, respectively, as derived from our proposed cosmological model. These results suggest that the inclusion of additional observational data significantly influences the location of the transition redshift, progressively shifting it to lower values when more precise constraints are introduced.

Furthermore, recent studies indicate that this transition redshift typically occurs within the redshift interval $0.2 \lesssim z \lesssim 2$ \cite{ref45,ref46}, marking a key epoch in the evolution of dark energy. Our findings are in good agreement with this observed range, particularly when utilizing the combined datasets OHD+PP and OHD+PP\&SH0ES. This consistency reinforces the reliability of our model in capturing the dynamical nature of dark energy and highlights its compatibility with recent observational trends.\\

\begin{figure}[hbt!]
    \centering
    \includegraphics[width=0.8\linewidth]{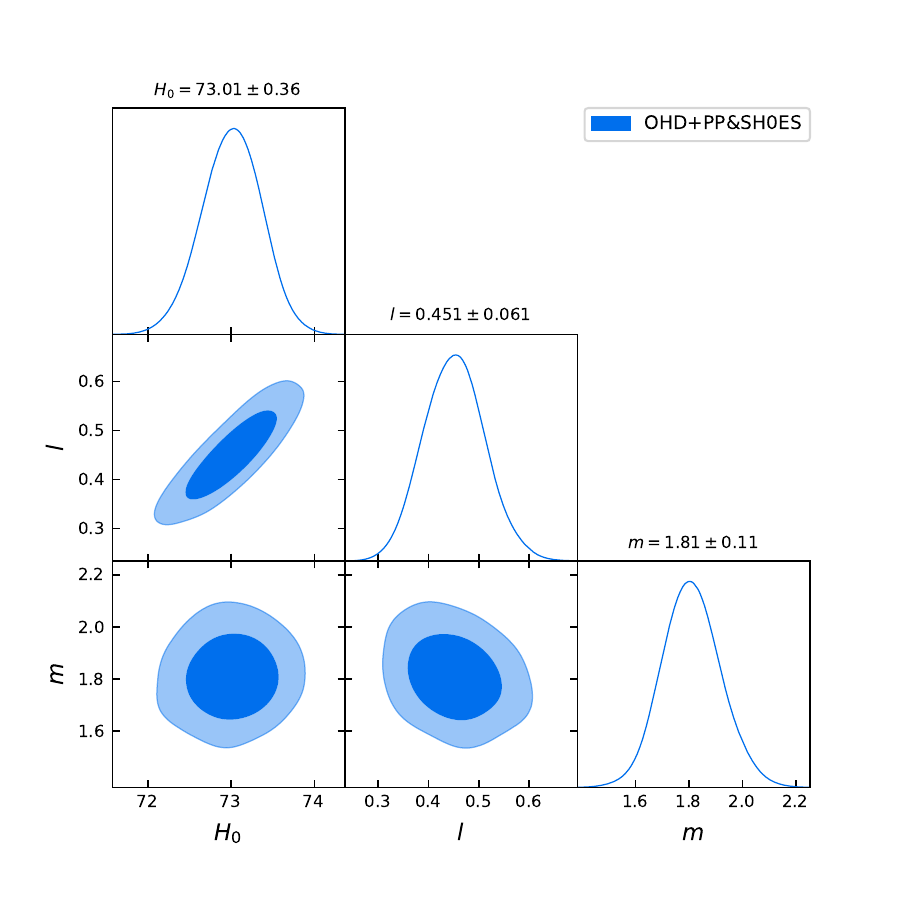}
   
    \caption{The triangular plot with $1\sigma$ and $2\sigma$ confidence levels for our model based on OHD+PP\&SH0ES datasets. }
    \label{fig6}
\end{figure}
In this context, the Hubble constant $H_0$ is one of the most fundamental parameters in cosmology, representing the current expansion rate of the universe. It plays a central role in determining the age, size, and evolution of the universe. However, despite significant advancements in observational techniques, a notable discrepancy has emerged between different methods of measuring $H_0$, leading to what is commonly referred to as the Hubble tension. This tension arises from the inconsistency between early-universe measurements—primarily those inferred from the Cosmic Microwave Background (CMB) by the Planck satellite under the assumption of the $\Lambda$CDM model—and late-universe measurements based on the local distance ladder, such as those from the SH0ES collaboration using supernovae calibrated by Cepheid variables are $5\sigma$. In the present work, we investigate the value of $H_0$ within the framework of the derived model using recently available observational datasets, including OHD, PP, and  PP\&SH0ES. Our analysis yields a constraint of \(H_0 = 72.12 \pm 2.1\) km s\(^{-1}\) Mpc\(^{-1}\) from OHD data sets. Also, we find the Hubble value \(H_0 = 72.73 \pm 0.31\) km s\(^{-1}\) Mpc\(^{-1}\) with a combination of OHD+PP data sets, and a similar value is obtained \(H_0 = 73.01 \pm 0.36\) km s\(^{-1}\) Mpc\(^{-1}\) by the joint analysis incorporating OHD+PP\&SH0ES datasets. Here, we compare our findings  with SH0ES collaboration's measurement of \(H_0 = 73.27 \pm 1.04\) km s\(^{-1}\) Mpc\(^{-1}\)  at the 68\% CL., based on supernovae calibrated with Cepheid variables \cite{ref47}.
We find that tensions $0.53\sigma$ (for the OHD+PP dataset) and $0.24\sigma$ (for the OHD+PP+SH0ES dataset) arise when comparing the Hubble constant derived from our model with the SH0ES value. However, our model successfully reduces the approximately $4.5\sigma$ Hubble tension that exists between the Planck \cite{ref48} and SH0ES measurements. We conclude that our model is consistent with the SH0ES calibration and supports a higher value of the Hubble constant. In this contrast, we observe from Fig.\ref{fig5} and Fig.\ref{fig6} that the parameter $H_0$ exhibits a positive correlation with the free parameter $l$, indicating that an increase in $H_0$ is associated with an increase in $l$. In both triangle plots, the parameters $H_0$, $l$, and $m$ are constrained using both 1D marginalized distributions and 2D joint confidence contours, based on the combined datasets OHD+PP and OHD+PP\&SH0ES.\\

\begin{figure}[hbt!]
    \centering
    \includegraphics[width=0.49\linewidth]{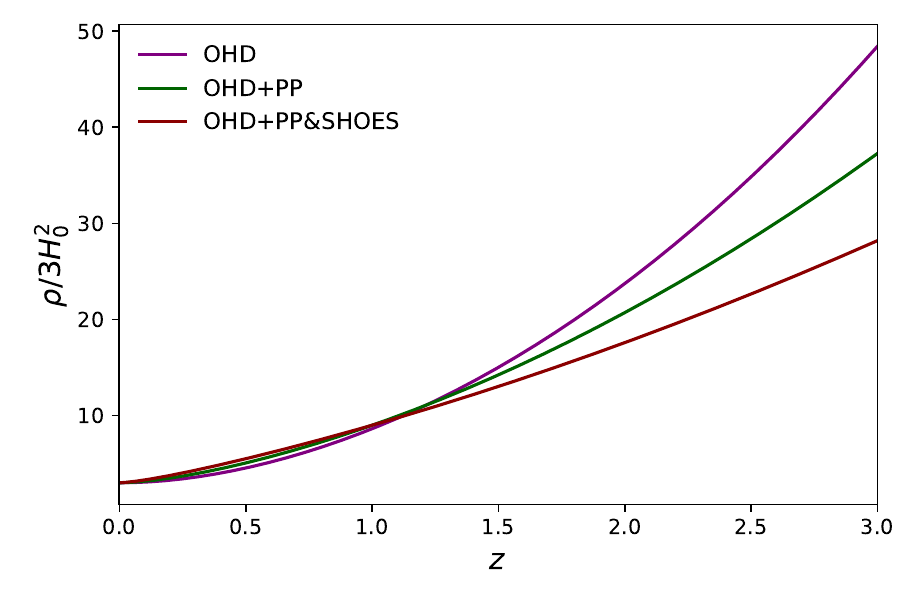}
    \includegraphics[width=0.49\linewidth]{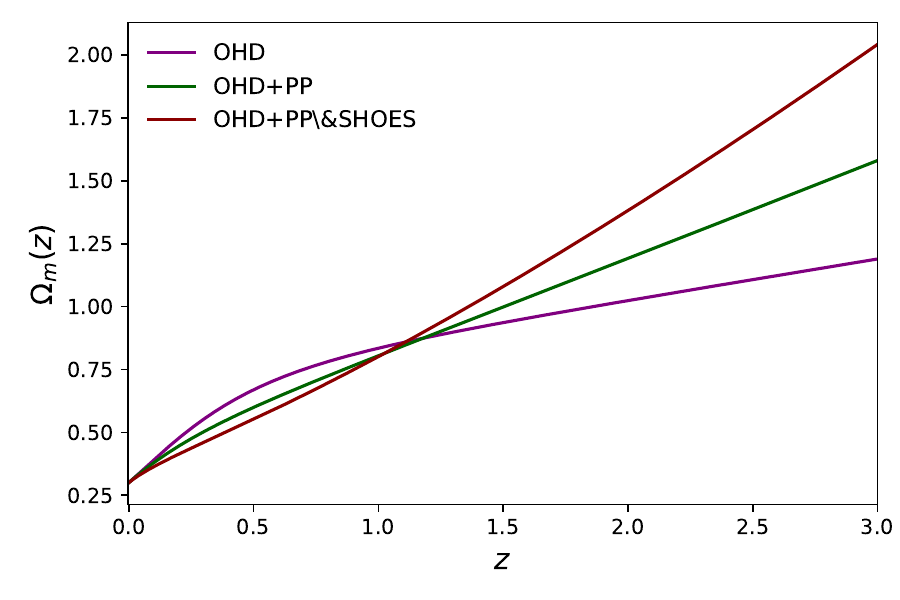}
    \caption{The reconstructed  trajectories (from present to past) of total density parameter and Matter density parameter in our model based on the given datasets.}
    \label{fig7}
\end{figure}

 The left panel of Fig.\ref{fig7} shows the expected trend of the total density parameter, which decreased as the universe expanded in the present epoch. It is important to note that the density parameter shown in this figure corresponds to the total matter-energy content of the universe, incorporating both dark matter and dark energy. Consequently, this parameter is expected to increase with redshift for all viable dark energy models, including the cosmological constant ($\Lambda$CDM), quintessence, and phantom scenarios, because the universe was more compact and denser at earlier times. Furthermore, the right panel of Fig.\ref{fig7} demonstrates the increasing trend of $\Omega_m(z)$ with redshift. This behavior is consistent with theoretical expectations and is observed across all the datasets considered within the framework of our derived model.\\

\begin{figure}[hbt!]
	    \centering
	    \includegraphics[width=0.49\linewidth]{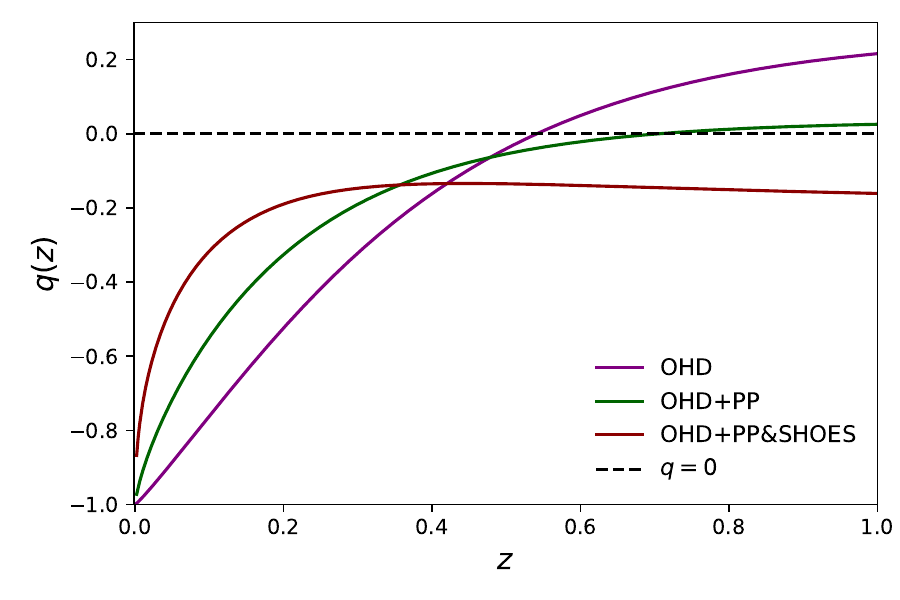}
        \includegraphics[width=0.49\linewidth]{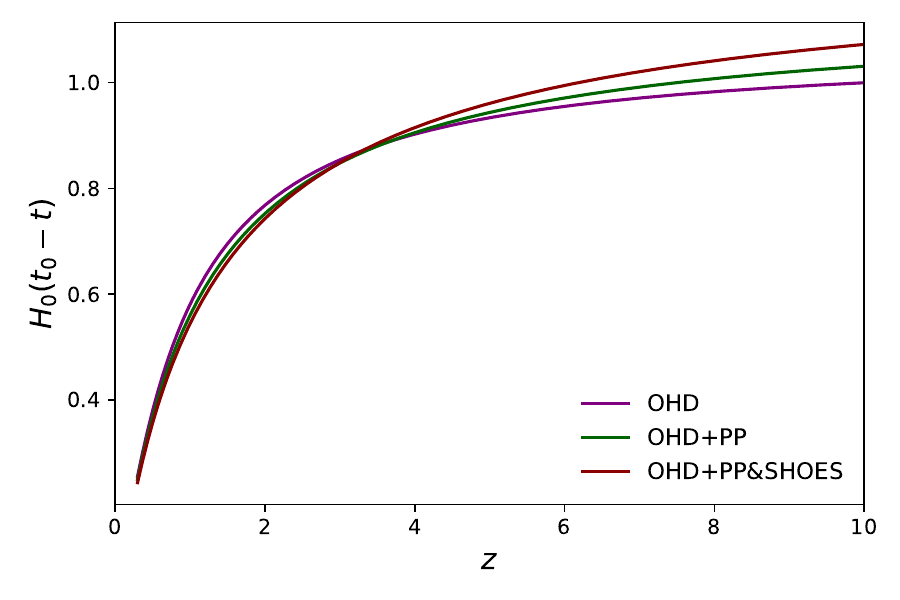}
	    \caption{The reconstructed  trajectories (from present to past) of deceleration parameter and and age of universe in our model based on the given datasets. }
	    \label{fig8}
	\end{figure}	
The deceleration parameter is a crucial parameter in cosmology that shows the behavior and evolution of the universe's expansion rate, represented by $q(z)$ and calculated using Eq.\ref{eq9}  as a function of the redshift for our derived model. A negative value ($q < 0$) denotes faster growth of the accelerating universe, whereas, a positive deceleration parameter ($q > 0$) indicates that the expansion rate of the universe shows down. A constant  expansion rate was observed if $q = 0$. This parameter plays a key role in defining the universe's final fate and offers important insights into its dynamics, particularly the impact of dark energy. In this study, we describe the characteristics of the deceleration parameter derived from $Om(z)$ diagnostic process.\\

In the left panel of Fig.\ref{fig8}, we present a 2D plot of $q(z)$ versus $z$ based on our derived model using various observational data sets such as OHD, OHD+PP, and OHD+PP\&SH0ES. The purple curve represents the results from the OHD dataset and shows an increasing trend, intersecting the black dotted line ($q = 0$) at a transition redshift of approximately $z_{\text{tr}} \approx 0.5$, indicating a shift from decelerated to accelerated expansion. When the PP dataset is included alongside OHD (green curve), the transition occurs at a higher redshift of approximately $z_{\text{tr}} \approx 0.75$. In contrast, the red curve—corresponding to the combined OHD+PP\&SH0ES data—remains entirely below the $q = 0$ line, indicating that the universe accelerates throughout the entire observed redshift range. These transition redshift values are derived from the $Om(z)$ diagnostic process from the OHD and OHD+PP data combinations, are in good agreement with recent results reported in the literature \cite{ref49,ref50,ref51,ref52,ref53}.\\

\textbf{Age of the Universe:} The age of the universe can be estimated using the lookback time, which represents the time interval between the present age of the universe, $t_0$, and its age at a given redshift $z$. In a cosmological model where the Hubble parameter $H(z)$ varies with the redshift, and the lookback time is calculated using the following integral.

We calculate the age of the universe as 
\begin{equation}
t_0 - t(z) = \int_0^z \frac{dz'}{(1 + z') H(z')}
\label{eq14}
\end{equation}

From  Eq.\ref{eq6} and Eq.\ref{eq14}, we get

\begin{equation}
H_0 (t_0 - t(z)) = \int_0^z \frac{dz'}{(1+z') \sqrt{1 + (1+z')^m . z'^{1 - l} \left(3 + 3z' + z'^2\right)}}
\label{eq15}
\end{equation}

 The beginning of the universe corresponds to \( t = 0 \) as \( z \to \infty \), and thus Eq.~\ref{eq15} reduces to
 
 \begin{equation}
H_0 t_0 = \int_0^{\infty} \frac{dz'}{(1+z') \sqrt{1 + (1+z')^m \cdot z'^{1 - l} \left(3 + 3z' + z'^2\right)}}
 \end{equation}

The right panel in Fig.\ref{fig8}  illustrates the variation in the normalized lookback time, denoted as $H_{0}(t_{0}-t)$, as a function of redshift 
$z$, for three different observational datasets: OHD (red curve), OHD+PP (green curve), and the combined dataset OHD+PP\&SH0ES (red curve). The lookback time $(t_{0}-t)$ represents the temporal separation between the present epoch and the cosmic time corresponding to a given redshift $z$, and is normalized here by the Hubble constant $H_{0}$ to render it dimensionless and suitable for comparative analysis.All three curves exhibit a monotonically increasing trend with redshift, consistent with cosmological expectations, wherein higher redshifts correspond to earlier epochs in the universe’s history. It is important to note
that the empirical value of the age of the Universe in Plank
collaboration results \cite{ref48} is obtained as $t_{0} =13.81 \pm 0.038$
Gyrs. In some other cosmological investigations, age of
the Universe is estimated as, $13.50\pm
0.23$ Gyrs \cite{ref53} and $13.20^{+3.6}_ {-2.0}$ Gyrs \cite{ref55}. In this paper, the present age of universe for the derived model is estimated as $t_{0} =13.21 \pm 1.02$  Gyrs for OHD, $t_{0} =13.51 \pm 0.53$ Gyrs for OHD+PP and $t_{0} =13.89 \pm 0.40$ Gyrs for OHD + PP$\&$ SH0ES.\\

\textbf{Statistical Analysis:}  
The minimum chi-square value, $\chi^2_{\min}$, obtained for each model indicates that the corresponding model provides an acceptable fit to the observational data as presented in Table \ref{tab1}. However, because the models considered involve different numbers of free parameters, a direct comparison based solely on $\chi^2_{\min}$ is not statistically meaningful. To address this issue, we employ standard information criteria for model comparison, namely the Akaike Information Criterion (AIC)~\cite{ref56} and the Bayesian (or Schwarz) Information Criterion (BIC)~\cite{ref57}.

\begin{table*}[hbt!]
	\caption{The difference, $\Delta \text{AIC} = \text{AIC}_{\text{our model}} - \text{AIC}_{\Lambda \text{CDM}}$ and $\Delta \text{BIC} = \text{BIC}_{\text{our model}} - \text{BIC}_{\Lambda \text{CDM}}$ for our model with respect to $\Lambda$CDM  from all considered data sets.}
	\label{tab1}
	\scalebox{0.85}{
		\begin{tabular}{lcccccc}
			\hline
			\toprule
			\textbf{Dataset }&\;\;\;\;\;\;\;\;\textbf{Model}\;\;\;\;\;\;\;\;& \textbf{AIC}\;& \;\;\;\;\;\;\;\textbf{$\triangle$AIC}\;\;\;\;\;
			\;\;\;\;&\textbf{BIC}\;\;\;&\;\; \textbf{$\triangle$BIC}\;\;\;\;\;\;\ & \textbf{$\chi^2$}\\
			\hline
			\textbf{CC} & \textbf{\text{Our model}}\,&\textbf{\text{34.23}},&\textbf{\text{2.17}}&\textbf{\text{38.71}}&\textbf{\text{2.18}}&\textbf{\text{28.23}}\vspace{0.2cm}\\
			&\textcolor{teal}{\textbf{$\bm{\Lambda}$CDM}} & \textcolor{teal}{\textbf{32.06}} & \textcolor{teal}{\textbf{-}}  & \textcolor{teal}{\textbf{36.54}} & \textcolor{teal}{\textbf{-}} & \textcolor{teal}{\textbf{26.06}}\vspace{0.1cm}\\
			\hline

			\textbf{CC+PP} & \textbf{\text{Our model}}\,&\textbf{\text{1455.56}}\,&\textbf{\text{3.85}}&\textbf{\text{1471.93}}&\textbf{\text{3.86}}&\textbf{\text{1449.56}}\vspace{0.2cm}\\
			&\textcolor{teal}{\textbf{$\bm{\Lambda}$CDM}}\ & \textcolor{teal}{\textbf{1451.71}}\, & \textcolor{teal}{\textbf{-}}  & \textcolor{teal}{\textbf{1468.08}} & \textcolor{teal}{\textbf{-}} & \textcolor{teal}{\textbf{1445.71}}
			 \vspace{0.1cm}\\ \hline

				\textbf{CC+PP\&SH0ES} & \textbf{\text{Our model}}\,&\textbf{\text{1356.81}}\,&\textbf{\text{2.70}}&\textbf{\text{1373.25}}&\textbf{\text{2.71}}&\textbf{\text{1350.81}}\vspace{0.2cm}\\
			&\textcolor{teal}{\textbf{$\bm{\Lambda}$CDM}}\ & \textcolor{teal}{\textbf{1354.11}}\, & \textcolor{teal}{\textbf{-}}  & \textcolor{teal}{\textbf{1370.55}} & \textcolor{teal}{\textbf{-}} & \textcolor{teal}{\textbf{1348.11}}\vspace{0.1cm}
			\\ \hline

		\end{tabular}
	}
\end{table*}

These criteria are widely used in statistical data analysis and introduce penalties for model complexity, thereby discouraging overfitting. The AIC and BIC are defined as
\begin{equation}
	\text{AIC} = -2 \ln \mathcal{L} + 2d = \chi^2_{\min} + 2d,
\end{equation}
\begin{equation}
	\text{BIC} = -2 \ln \mathcal{L} + d \ln N = \chi^2_{\min} + d \ln N,
\end{equation}
where the maximum likelihood is given by $\mathcal{L} = \exp(-\chi^2_{\min}/2)$, $d$ denotes the number of free parameters in the model, and $N$ denotes the total number of data points used in the analysis.\\

For a meaningful comparison, we adopted standard $\Lambda$CDM cosmology as the reference (baseline) model. The relative performance of any alternative model $Q$ is quantified by the difference $\Delta X = X_Q - X_{\Lambda\mathrm{CDM}}$, where $X$ represents either the AIC or BIC. The level of support for a given model can be interpreted as follows: For  AIC, strong support corresponds to $\Delta \mathrm{AIC} \leq 2$, moderate support corresponds to $4 \leq \Delta \mathrm{AIC} \leq 7$, and essentially no support for $\Delta \mathrm{AIC} \geq 10$. For  BIC, the evidence is considered positive if $2 \leq \Delta \mathrm{BIC} \leq 6$, strong if $6 < \Delta \mathrm{BIC} \leq 10$, and \emph{very strong} if $\Delta \mathrm{BIC} > 10$.\\

Table~\ref{tab1} summarizes the computed values of $\chi^2$, AIC, and BIC for our model in comparison with the standard $\Lambda$CDM cosmology, using three different observational datasets: OHD, combined OHD+PP, and OHD+PP+SH0ES compilations. For a meaningful comparison, we also present the relative differences $\Delta$AIC and $\Delta$BIC with respect to the reference $\Lambda$CDM model. We find that the $\Delta$BIC values lie within the range $2 \leq \Delta \mathrm{BIC} \leq 6$, indicating that our model is consistent with both the observational datasets and the standard cosmological model from the OHD, OHD+PP, and OHD+PP+SH0ES datasets. Similarly, the $\Delta$AIC values are close to lies $(2,4)$, suggesting that our model and the $\Lambda$CDM model provide comparably good fit to the data. Overall, these results demonstrate that our model remains statistically consistent and is in agreement with observational constraints.

\color{black}

\section{Conclusion}

In this work, we presented a new parametric approach to the \$Om(z)\$ diagnostic to explore the dynamic behavior of dark energy and  its implications on the expansion history of the universe. Our model, characterized by two free parameters, $l$ and $m$, successfully captures a wide range of dark energy behaviors, ranging from quintessence-like to phantom-like regimes, by appropriately tuning the parameter values. 
The generalized form of ${Om(z)}$, given by ${Om(z)} = \dfrac{z^l}{(1+z)^m}$, was employed to derive expressions for key cosmological quantities, including the Hubble parameter $H(z)$, total energy density ${\rho_{\text{tot}}(z)}$, equation of state parameter $\textcolor{blue}{w_{\text{tot}}(z)}$, 
and the deceleration parameter $q(z)$. These analytical forms enabled us 
to perform a detailed investigation of the univers's evolution 
using various observational datasets. Using recent Observational Hubble Data (OHD), PantheonPlus (PP), and PantheonPlus $\&$ SH0ES (PP\&SH0ES), we constrained our model parameters using MCMC techniques. 
Our analysis shows that the $O{m}(z)$ diagnosis based on this parametrization is highly flexible and consistent with current cosmological observations. A key finding is the identification of transition redshifts—where the nature of dark energy shifts from a phantom to a quintessence-like regime as $z = 1.31$ (OHD), $z = 0.65$ (OHD+PP), and $z = 0.33$ (OHD+PP\&SH0ES). These values are consistent with those in the literature and suggest that the inclusion of additional data leads to an earlier transition epoch, underscoring the importance of combining independent observational probes. A Planck Collaboration's analysis of CMB data within the $\Lambda$CDM framework yields $H_{0} = (67.4 \pm 0.5) km s^{-1} MPc^{-1}$ at 68\% confidence level (CL) \cite{NA:2020}. In contrast, the SH0ES Collaboration (Supernovae $H_{0}$ for the Equation of State ) found a significantly higher value of $H_{0} = (73.04 \pm 1.04) km s^{-1} MPc^{-1}$ at 68\% CL, using the three-rung distance ladder method with Cepheids \cite{ref43a}. A more comprehensive review addressing Hubble tension can be found in \cite{R11,R12,R13,R14,R15}. Furthermore, our model provides updated constraints on the Hubble constant,  yielding \(H_0 = 72.12 \pm 2.1\) km s\(^{-1}\) Mpc\(^{-1}\) (OHD), \(H_0 = 72.73 \pm 0.31\) km s\(^{-1}\) Mpc\(^{-1}\) (OHD+PP), and  \(H_0 = 73.01 \pm 0.36\) km s\(^{-1}\) Mpc\(^{-1}\) (OHD+PP$\&$SH0ES). These values lie in close agreement with the local SH0ES measurement of  \(H_0 = 73.27 \pm 1.04\) km s\(^{-1}\) Mpc\(^{-1}\)  and effectively alleviate the long standing Hubble tension with Planck CMB estimates. The reduction in tension-down to $0.24\sigma$ in the joint dataset-indicates that our model offers a viable alternative to the standard $\Lambda$CDM framework in reconciling early and late-time measurements of the universe's expansion rate. Our model also estimates the present age of the Universe as $t_0 = 13.21 \pm 1.02$ Gyr (OHD), $t_0 = 13.51 \pm 0.53 $ Gyr (OHD+PP), and $t_0 = 13.89 \pm 0.40$ Gyr (OHD+PP\&SH0ES), values that are in good agreement with both Planck results and other recent astrophysical estimates. These findings confirm the robustness of the proposed \$Om(z)\$ parameterization for characterizing the expansion history and age of the universe.\\

In summary, our analysis demonstrates that the proposed $Om(z)$ model not only offers a unified framework for analyzing dark energy dynamics but also provides competitive fits to current observational data. It effectively accommodates the transition between dark-energy regimes, mitigates the Hubble tension, and yields a consistent cosmic age. Future work could involve extending this framework to include additional cosmological probes such as BAO and CMB data, or testing its predictions within the context of modified gravity theories.

\section*{Declaration of competing interest}
The authors declare that they have no known competing financial
interests or personal relationships that could have influenced
the work reported in this study.

\section*{Data availability}
We employe the  publicly available Pantheon Plus (PP) data and Observational Hubble Parameter (OHD) data presented in this study. The OHD data are compiled from publicly available cosmic chronometer measurements in the literature, with a representative compilation accessible at: \href{https://github.com/AhmadMehrabi/Cosmic_chronometer_data}{https://github.com/AhmadMehrabi/Cosmic chronometer data}.The Pantheon Plus (PP) compilation (distance moduli and covariance matrices), which is publicly available on GitHub: \href{https://github.com/brinckmann/montepython_public/tree/3.6/montepython/likelihoods/Pantheon_Plus}{https://github.com/brinckmann/montepython public/tree/3.6/montepython/likelihoods/Pantheon Plus}. No additional data were used in this study.

\begin{acknowledgments}
\noindent 
The authors (AD \& AP) are thankful to Inter-University Centre for Astronomy \& Astrophysics (IUCAA), Pune, India for providing support and facility under Visiting
Associateship program. M. Yadav was sponsored by a senior Research Fellowship (CSIR Ref. No. 180010603050) from the Council of Scientific and Industrial Research,  Government of India.
The authors thank the anonymous reviewer and esteemed Editor for their helpful remarks, which improved the quality of the manuscript in its present form.  

\end{acknowledgments}


\begin{thebibliography}{99} 


\bibitem{ref1}
A.~G.~Riess \textit{et al}., Observational evidence from supernovae for an accelerating universe and a cosmological constant, Astron. J. \textbf{116}, 1009 (1998).


\bibitem{ref2}
{S. Perlmutter, G. Aldering, \textit{et al}., Measurements of $\Omega$ \text{and } $\Lambda$ from 42 high-redshift supernovae, Astrophys. J. \textbf{517}, 565 (1999)}.


\bibitem{ref3}
A. G. Riess \textbf{et al}., New Hubble space telescope discoveries of type Ia supernovae at $z \geq 1$: narrowing constraints on the
early behavior of dark energy, Astron. Astrophys. \textbf{659}, 98 (2007).


\bibitem{ref4}
G. F. Smoot, C. L. Bennett, \textit{et al}., Structure in the COBE differential microwave radiometer first-year maps, Astron.
Astrophys. \textbf{396}, L1 (1992).

\bibitem{ref5}
C. L. Bennett, A. J. Banday, \textit{et al}., Four-year COBE DMR cosmic microwave background observations: maps and basic
results, Astrophys. J. Lett.\textbf{ 464}, L1 (1996).

\bibitem{ref6}
D. N. Spergel, L. Verde, \textit{et al}., First-year Wilkinson Microwave Anisotropy Probe (WMAP) observations: determination
of cosmological parameters, Astrophys. J. Suppl. \textbf{148}, 175 (2003).

\bibitem{ref7}
T. Koivisto and D. F. Mota, Dark energy anisotropic stress and large scale structure formation, Phys. Rev. D \textbf{73}, 083502 (2006).


\bibitem{ref8}
D. J. Eisenstein, I. Zehavi, \textit{et al}., Detection of the baryon acoustic peak in the large-scale correlation function of SDSS
luminous red galaxies, Astron. Astrophys. \textbf{633}, 560 (2005).

\bibitem{ref9}
C. Blake, E. A. Kazin, \textit{et al}., The WiggleZ Dark Energy Survey: mapping the distance–redshift relation with baryon
acoustic oscillations, Mon. Not. R. Astron. Soc. \textbf{418}, 1707 (2011).

\bibitem{ref10}
F. Beutler, C. Blake, \textit{et al}., The 6dF Galaxy Survey: baryon acoustic oscillations and the local Hubble constant, Mon.
Not. R. Astron. Soc. \textbf{416}, 3017 (2011).
\bibitem{ref11}
A Dixit, A Pradhan, K Ghaderi,Interacting Bianchi Type-V Universe: Observational Constraints, Gravitation and Cosmology {\bf 30} 376-391 (2024).

\bibitem{ref12}
S. Weinberg, The cosmological constant problem, Rev. Mod. Phys. \textbf{61}, 1 (1989).


\bibitem{ref13}
P. J. Steinhardt, L. Wang, and I. Zlatev, Cosmological tracking solutions, Phys. Rev. D \textbf{59}, 123504 (1999).

\bibitem{ref14}
B. Ratra and P. J. E. Peebles, Cosmological consequences of a rolling homogeneous scalar field, Phys. Rev. D \textbf{37}, 3406 (1988).

\bibitem{ref15}
T. Chiba, T. Okabe, and M. Yamaguchi, Kinetically driven quintessence, Phys. Rev. D \textbf{62}, 023511 (2000).

\bibitem{ref16}
C. Armendariz-Picon, V. Mukhanov, and P. J. Steinhardt, Dynamical solution to the problem of a small cosmological constant and late-time cosmic acceleration, Phys. Rev. Lett. \textbf{85}, 4438 (2000).


\bibitem{ref17}
M. Sami and A. Toporensky, Phantom field and the fate of the universe, Mod. Phys. Lett. A \textbf{19}, 1509 (2004).

\bibitem{ref18}
M. Sami, A. Toporensky, P. V. Tretjakov, and S. Tsujikawa, The fate of (phantom) dark energy universe with string curvature corrections, Phys. Lett. B \textbf{619}, 193 (2005).

\bibitem{ref19}
W. Yang, A. Mukherjee, E. Di Valentino, and S. Pan, Interacting dark energy with time varying equation of state and the $H_0$ tension, Phys. Rev. D \textbf{98}, 123527 (2018).


\bibitem{ref20}
Y. Tada and T. Terada, Quintessential interpretation of the evolving dark energy in light of DESI observations, Phys. Rev. D \textbf{109}, L121305 (2024).

\bibitem{ref21}
L. A. Escamilla, W. Giarè, E. Di Valentino, R. C. Nunes, and S. Vagnozzi, The state of the dark energy equation of state circa 2023, JCAP \textbf{2024}(05), 091 (2024).

\bibitem{ref22}
I. D. Gialamas, G. H\"utsi, K. Kannike, A. Racioppi, M. Raidal, M. Vasar, and H. Veerm\"ae, Interpreting DESI 2024 BAO: late-time dynamical dark energy or a local effect?, Phys. Rev. D \textbf{111}, 043540 (2025).

%%%%%%%%%%%%%%%%%%%%%%%%%%%%%%%%%%%%%%%%%%%%%%%%%%%%%%%%%%%%%%%%%%%%%%%%%%%%%%%%%%%

\bibitem{ref23}
V. Sahni, A. Shafieloo, and A. A. Starobinsky, Two new diagnostics of dark energy, Phys. Rev. D \textbf{78}, 103502 (2008).

\bibitem{ref24}
V. Sahni, A. Shafieloo, and A. A. Starobinsky, Model-independent evidence for dark energy evolution from baryon acoustic oscillations, Astrophys. J. \textbf{793}, L40 (2014).

\bibitem{ref25}
X. Ding, M. Biesiada, S. Cao, Z. Li, and Z.-H. Zhu, Is there evidence for dark energy evolution?, Astrophys. J. Lett. \textbf{803}, L22 (2015).

\bibitem{ref26}
X. Zheng, X. Ding, M. Biesiada, S. Cao, and Z.-H. Zhu, What are the Om$h^2(z_1, z_2)$ and Om$(z_1, z_2)$ diagnostics telling us in light of $H(z)$ data?, Astrophys. J. \textbf{825}, 17 (2016).

\bibitem{ref27}
M. Seikel, S. Yahya, R. Maartens, and C. Clarkson, Using $H(z)$ data as a probe of the concordance model, Phys. Rev. D \textbf{86}, 083001 (2012).

\bibitem{ref28}
N. Myrzakulov, M. Koussour, and D. J. Gogoi, A new Om$(z)$ diagnostic of dark energy in general relativity theory, Eur. Phys. J. C \textbf{83}, 594 (2023).

\bibitem{ref29}
Y. Myrzakulov, A. H. A. Alfedeel, M. Koussour, E. I. Hassan, and S. Muminov, Diagnostics of dark energy evolution using logarithmic Om$(z)$ parameterization, J. High Energy Astrophys. \textbf{47}, 100386 (2025).

\bibitem{ref30}
J.-Z. Qi, S. Cao, M. Biesiada, T.-P. Xu, Y. Wu, S.-X. Zhang, and Z.-H. Zhu, What do parameterized Om$(z)$ diagnostics tell us in light of recent observations?, Res. Astron. Astrophys. \textbf{18}, 066 (2018).

\bibitem{ref31}
A. Shafieloo, V. Sahni, and A. A. Starobinsky, New null diagnostic customized for reconstructing the properties of dark energy from baryon acoustic oscillations data, Phys. Rev. D \textbf{86}, 103527 (2012).

\bibitem{ref32}
S. Yahya, M. Seikel, C. Clarkson, R. Maartens, and M. Smith, Null tests of the cosmological constant using supernovae, Phys. Rev. D \textbf{89}, 023503 (2014).

\bibitem{ref33}
C. Zhang, H. Zhang, et \textit{al}., Four new observational $H(z)$ data from luminous red galaxies in the Sloan Digital Sky Survey
data release seven, Res. Astron. Astrophys. \textbf{14}, 1221 (2014).

\bibitem{ref34}
J. Simon, L. Verde, and R. Jimenez, Constraints on the redshift dependence of the dark energy potential,Phys. Rev. D
\textbf{71}, 123001 (2005).

\bibitem{ref35}
M. Moresco \textit{et al}., Improved constraints on the expansion rate of the Universe up to $z \sim 1.1$ from the spectroscopic
evolution of cosmic chronometers, JCAP \textbf{08}, 006 (2012).

\bibitem{ref36}
M. Moresco, L. Pozzetti, \textit{et al}., A 6\% measurement of the Hubble parameter at $z \sim 0.45$: direct evidence of the epoch of
cosmic re-acceleration," JCAP \textbf{05}, 014 (2016).

\bibitem{ref36a}

{D. Stern, R. Jimenez, \textit{et al}., Cosmic chronometers: constraining the equation of state of dark energy. I: H (z) measurements, JCAP \textbf{02}, 008 (2010).}

\bibitem{ref37}
A. L. Ratsimbazafy, S. I. Loubser, S. M. Crawford, \textit{et al}., Age-dating Luminous Red Galaxies observed with the Southern
African Large Telescope, Mon. Not. Roy. Astron. Soc. \textbf{467}, 3239 (2017).

\bibitem{ref38}
N. Borghi, M. Moresco, and A. Cimatti, Toward a Better Understanding of Cosmic Chronometers: A New Measurement of
$H(z) \rm{at} z \sim 0.7$, Astrophys. J. Lett. \textbf{928}, L4 (2022).

\bibitem{ref39}
K. Jiao, N. Borghi, M. Moresco, and T.-J. Zhang, New Observational $H(z)$ Data from Full-spectrum Fitting of Cosmic
Chronometers in the LEGA-C Survey, Astrophys. J. Suppl. \textbf{265}, 48 (2023).

\bibitem{ref40}
M. Moresco, Raising the bar: new constraints on the Hubble parameter with cosmic chronometers at $z \sim 2$, Mon. Not.
Roy. Astron. Soc. \textbf{450}, L16 (2015).



\bibitem{ref41}
R. Jimenez and A. Loeb, Constraining cosmological parameters based on relative galaxy ages, Astrophys. J. \textbf{573}, 37 (2002).

\bibitem{ref42}
{D. M. Scolnic \textit{et al}., The Complete Light-curve Sample of Spectroscopically Confirmed SNe Ia from Pan-STARRS1 and Cosmological Constraints from the Combined Pantheon Sample, Astrophys. J. \textbf{859}, 101 (2018)}.   

\bibitem{ref43}
{D. Scolnic, D. Brout, \textit{et al}., The Pantheon+ Analysis: The Full Data Set and Light-curve Release, Astrophys. J. \textbf{938}, 113 (2022)}.

\bibitem{ref43a}
{A.G Riess \textit{et al}, A Comprehensive Measurement of the Local Value of the Hubble Constant with  Uncertainty from the Hubble Space Telescope and the SH0ES Team, Astrophys. J. Lett. \textbf{934}, L7 (2022)}.  

\bibitem{ref44}
D. F. Mackey \textit{et al}.,emcee: the MCMC hammer ,Publ. Astron. Soc. Pac. \textbf{125}, 306 (2013).

\bibitem{ref45}
M. Scherer, M. A. Sabogal, R. C. Nunes, and A. De Felice, Challenging $\Lambda$CDM: $5\sigma$ evidence for a dynamical dark energy late-time transition, arXiv:2504.20664 [astro-ph.CO] (2025).

\bibitem{ref46}
\"O. Akarsu, L. Perivolaropoulos, A. Tsikoundoura, A. E. Y\"ukselci, and A. Zhuk, Dynamical dark energy with AdS-to-dS and dS-to-dS transitions: Implications for the $H_0$ tension, arXiv:2502.14667 [astro-ph.CO] (2025).

\bibitem{ref47}
D. Brout, D. Scolnic, \textit{et al}., The Pantheon+ analysis: cosmological constraints, Astrophys. J. \textbf{938}, 110 (2022).

\bibitem{ref48}
N. Aghanim \textit{et al}. (Planck), Planck 2018 results. VI. Cosmological parameters, Astron. Astrophys. \textbf{641}, A6 (2020).

\bibitem{ref49}
O. Farooq, F. R. Madiyar, S. Crandall, and B. Ratra, Hubble parameter measurement constraints on the redshift of the deceleration-acceleration transition, dynamical dark energy, and space curvature, Astrophys. J. \textbf{835}, 26 (2017).

\bibitem{ref50}
J. Rom\'an-Garza, T. Verdugo, J. Maga\~na, and V. Motta, Constraints on barotropic dark energy models by a new phenomenological $q(z)$ parameterization, Eur. Phys. J. C \textbf{79}, 890 (2019).

\bibitem{ref51}
{A. Pradhan, D. C. Maurya, and A. Dixit, Dark energy nature of viscus universe in $f(Q)$-gravity with observational constraints, Int. J. Geom. Methods Mod. Phys. \textbf{18}, 2150124 (2021)}.

\bibitem{ref52}
V. K. Bhardwaj, A. Dixit, R. Rani, G. K. Goswami, and A. Pradhan, An axially symmetric transitioning model with observational constraints, Chin. J. Phys. \textbf{80}, 261--274 (2022).

\bibitem{ref53}
{D. Valcin, R. Jimenez, L. Verde, J. L. Bernal, and B. D. Wandelt, The age of the Universe with globular clusters: reducing systematic uncertainties, JCAP \textbf{08}, 017 (2021).}
%%%%%%%%%%%%%%%%%%%%%%%%%%%%%%%%%%%%%%%%%%%%%%%%%%%%%%%%%%%%%%%%%%%%%

%\bibitem{ref54}
%A. Pradhan, A. Dixit, and D. C. Maurya, Quintessence behavior of an anisotropic bulk viscous cosmological model in modified $f(Q)$-gravity, Symmetry \textbf{14}, 2630 (2022).



\bibitem{ref55}
I. Ferreras, A. Melchiorri, and J. Silk, Setting new constraints on the age of the Universe, Mon. Not. R. Astron. Soc. \textbf{327}, L47--L51 (2001).

\bibitem{ref56}
H. Akaike, A new look at the statistical model identification, IEEE transactions on automatic control \textbf{19}, 716 (1974).

\bibitem{ref57}
G. E. Schwarz, Estimating the dimension of a model, Ann. Statist. {\bf 6}, 461 (1978).
\bibitem{NA:2020}
N. Aghanim, et al. (Planck Collaboration), Planck 2018 results. VI. Cosmological parameters, Astron. Astrophys. {\bf 641}, A6 (2020), arXiv:1807.06209 [astro-ph.CO].
\bibitem{R11}
E. Di Valentino, O. Mena, S. Pan, et al., In the realm of the Hubble tension-a review of solutions, Class. Quantum Grav. {\bf 38}, 153001 (2021), arXiv:2103.01183 [astro-ph.CO].		
\bibitem{R12}
M. Moresco, L. Amati, L. Amendola, et al., Unveiling the Universe with emerging cosmological probes, Living Rev. Relativ. {\bf 25}, 6 (2022). 
\bibitem{R13}
E. M$\ddot{o}$rtsell, A. Goobar, J. Johansson, and S. Dhawan, Sensitivity of the Hubble Constant Determination to Cepheid Calibration, ApJ {\bf 933}, 212 (2022).
\bibitem{R14}
L. Perivolaropoulos, Hubble tension or distance ladder crisis?, Phys. Rev. D {\bf 110}, 123518 (2024).
\bibitem{R15}
T. Liu, S. Cao, J. Wang, Probing potential redshift-dependent systematics in the Hubble tension: Model-independent $H_{0}$ constraints from DESI $R2$, Phys. Rev. D {\bf 112}, 123539 (2025). 

\end{thebibliography}
\end{document}